\documentclass[
  journal=pasa,
  manuscript=research-paper, 
  year=2026, 
  volume=YY,
]{cup-journal} 

\usepackage{amssymb,microtype,siunitx,booktabs}

\usepackage{tocloft}
\usepackage{lineno}
\usepackage{setspace}

\usepackage{tipa}

\usepackage{academicons} 
\usepackage{fontawesome5}

\definecolor{color1}{HTML}{1f77b4}
\definecolor{color2}{HTML}{ff7f0e}
\definecolor{color3}{HTML}{2ca02c}
\definecolor{color4}{HTML}{d62728}

\usepackage{amsmath,amsfonts}
\usepackage{graphicx}
\usepackage{relsize}

\usepackage{listings}
\usepackage{multirow}

\usepackage{longtable}
\usepackage{acro}
\usepackage{hyperref}
\hypersetup{
    colorlinks,
    linkcolor={color1},
    citecolor={color1},
    urlcolor={color1}
}
\usepackage{cleveref}
\usepackage{xspace}

\acsetup{
  single    = false,
  list/sort  = true,
  cite/group = true,
  cite/group/cmd = \citealt,
  cite/group/pre = {; \xspace},
  patch/longtable=false,
}

\newcommand{\amigo}{\textsc{Amigo}\xspace}
\newcommand{\camino}{\texttt{camino}\xspace}
\newcommand\equinox{\texttt{equinox}\xspace}

\newcommand\optax{\texttt{optax}\xspace}
\newcommand\numpyro{\texttt{numpyro}\xspace}
\newcommand\dlux{\textsc{$\partial$Lux}\xspace}
\newcommand\jax{\textsc{Jax}\xspace}

\DeclareAcronym{amigo}{
  short = \amigo,
  long = Aperture Masking Interferometry Generative Observations
}
\DeclareAcronym{psf}{
  short = PSF,
  long = Point Spread Function
}
\DeclareAcronym{otf}{
  short = OTF,
  long = Optical Transfer Function
}
\DeclareAcronym{ifu}{
  short = IFU,
  long = Integral Field Unit
}
\DeclareAcronym{fft}{
  short = FFT,
  long = Fast Fourier Transform
}
\DeclareAcronym{rom}{
  short = ROM,
  long = Reduced Order Modelling,
  cite = {Benner2015}
}
\DeclareAcronym{autodiff}{
  short = autodiff,
  long = Automatic Differentiation,
  cite={autodiff}
}
\DeclareAcronym{jwst}{
  short = JWST,
  long = James Webb Space Telescope,
  cite = {2006SSRGardner,2023PASPGardner}
}
\DeclareAcronym{niriss}{
  short = NIRISS,
  long = Near Infrared Imager and Slitless Spectrograph
}
\DeclareAcronym{nir}{
  short = NIR,
  long = Near Infrared
}
\DeclareAcronym{ami}{
  short = AMI,
  long = Aperture Masking Interferometer
}
\DeclareAcronym{bfe}{
  short = BFE,
  long = Brighter-Fatter Effect
}
\DeclareAcronym{edm}{
  short = EDM,
  long = Effective Detector Model
}
\DeclareAcronym{nn}{
  short = NN,
  long = Neural Network
}
\DeclareAcronym{cnn}{
  short = CNN,
  long = Convolutional Neural Network
}
\DeclareAcronym{ml}{
  short = ML,
  long = Machine Learning
}
\DeclareAcronym{fov}{
  short = FOV,
  long = field of view
}

\DeclareAcronym{h2rg}{
  short = H2RG,
  long = Teledyne HAWAII-2RG
}
\DeclareAcronym{adc}{
  short = ADC,
  long = Analogue to Digital Converter
}
\DeclareAcronym{ipc}{
  short = IPC,
  long = Inter-Pixel Capacitance
}
\DeclareAcronym{disco}{
  short = DISCO,
  long = Delay-Insensitive Subspace of Calibrated Observables
}
\DeclareAcronym{nrm}{
  short = NRM,
  long = Non-Redundant Mask
}
\DeclareAcronym{hst}{
  short = HST,
  long = \textit{Hubble Space Telescope}
}
\DeclareAcronym{mcmc}{
  short = MCMC,
  long = Monte Carlo Markov Chain,
  cite={Metropolis1953}
}
\DeclareAcronym{iwa}{
  short = IWA,
  long = Inner Working Angle
}
\DeclareAcronym{opd}{
  short = OPD,
  long = Optical Path Difference
}
\DeclareAcronym{go}{
  short = GO,
  long = General Observer
}
\DeclareAcronym{sgd}{
  short = SGD,
  long = stochastic gradient descent,
  cite= {Ruder2016}
}
\DeclareAcronym{hmc}{
  short = HMC,
  long = Hamiltonian Monte Carlo,
  cite= {Betancourt2017}
}
\DeclareAcronym{mast}{
  short = MAST,
  long = Mikulski Archive for Space Telescopes
}

\DeclareAcronym{gpu}{
  short = GPU,
  long = Graphics Processing Unit
}

\DeclareAcronym{wfs}{
  short = WFS\&C,
  long = Wavefront Sensing and Control
}

\DeclareAcronym{lct}{
  short = LCT,
  long = Linear Canonical Transform
}

\title[Four Years of JWST/NIRCam Wavefronts]{Model-Based Retrieval of Four Years of JWST/NIRCam Wavefronts}

\author{\href{https://orcid.org/0000-0003-2259-3911}{Shrishmoy~Ray}}
\affiliation{School of Mathematical and Physical Sciences, Macquarie University, Sydney, NSW 2109, Australia}
\alsoaffiliation{Astrophysics and Space Technologies Research Centre, Macquarie University, Sydney, NSW 2109, Australia}
\email[Shrishmoy~Ray]{shrishmoy.ray@mq.edu.au}

\author{\href{https://orcid.org/0000-0003-2595-9114}{Benjamin~Pope}}
\affiliation{School of Mathematical and Physical Sciences, Macquarie University, Sydney, NSW 2109, Australia}
\alsoaffiliation{Astrophysics and Space Technologies Research Centre, Macquarie University, Sydney, NSW 2109, Australia}

\author{\href{https://orcid.org/0000-0002-1015-9029}{Louis~Desdoigts}}

\affiliation{Leiden Observatory, Niels Bohrweg 2, Leiden 2300RA, The Netherlands}

\author{\href{https://orcid.org/0009-0008-8188-8668}{Hayden~Greer}}
\affiliation{Sydney Institute for Astronomy, School of Physics, The University of Sydney, NSW 2006, Australia}

\author{\href{https://orcid.org/0009-0000-9575-6190}{Milo~Langker}}
\affiliation{Sydney Institute for Astronomy, School of Physics, The University of Sydney, NSW 2006, Australia}

\author{\href{https://orcid.org/0000-0001-7026-6291}{Peter~Tuthill}}
\affiliation{Sydney Institute for Astronomy, School of Physics, The University of Sydney, NSW 2006, Australia}

\author{\href{https://orcid.org/0000-0002-3191-8151}{Marshall~Perrin}}
\affiliation{Space Telescope Science Institute, 3700 San Martin Drive, Baltimore, MD 21218, USA}

\author{\href{https://orcid.org/0000-0003-1251-4124}{Anand~Sivaramakrishnan}}
\affiliation{Space Telescope Science Institute, 3700 San Martin Drive, Baltimore, MD 21218, USA}
\alsoaffiliation{Astrophysics Department, American Museum of Natural History, 79th ST at CPW, New York, NY, 10024, USA}
\alsoaffiliation{Department of Physics and Astronomy, Johns Hopkins University, 3701 San Martin Drive, Baltimore, MD 21218, USA}

\author{\href{https://orcid.org/0000-0002-7162-8036}{\mbox{Alexandra~Greenbaum}}}
\affiliation{IPAC, Caltech, 1200 E. California Blvd., Pasadena, CA 91125, USA}

\author{\href{https://orcid.org/0000-0003-3818-408X}{Laurent~Pueyo}}
\affiliation{Space Telescope Science Institute, 3700 San Martin Drive, Baltimore, MD 21218, USA}

\author{\href{https://orcid.org/0009-0003-5950-4828}{Max~Charles}}
\affiliation{Sydney Institute for Astronomy, School of Physics, The University of Sydney, NSW 2006, Australia}

\received {30 Sep 2026}
\revised  {27 Oct 2026}
\accepted {30 Oct 2026}
\published{XX XXX 2026}


\begin{document}

\begin{abstract}

The fidelity of data recovered by the \textit{James Webb Space Telescope} (JWST) depends on precise monitoring of the mirror segments' surfaces and alignment. To this end, every few days the observatory takes pairs of defocused images of a reference star with its NIRCam instrument; a closed-source phase retrieval package is then employed to infer the wavefront. We here present \camino (Computational Aberration Modelling and Inference for NIRCam Observations), an open-source algorithm to model NIRCam built in \jax using the differentiable Fresnel optics code \dlux and regularised maximum likelihood inference that has been conditioned on the entire archive of NIRCam wavefront sensing data. We typically improve the reduced chi-squared in fits to the defocused data from $\chi^2_\nu{\sim}60{-}65$ to $\chi^2_\nu{\sim}12{-}13$. We reconstruct the wavefront history of JWST  from the beginning of science operations to early 2026. This open source implementation operates in parallel to the current STScI phase retrieval algorithm, allowing access by the broader scientific community to high-accuracy wavefront estimation and enabling its direct integration into science-analysis workflows. This represents a major step towards the ultimate goal of predictive calibration methods based on physics-based forward modelling, obviating the need for time-consuming observation of contemporaneous dedicated reference stars.

\end{abstract}

\maketitle
\section{Introduction}
\label{sec:intro}

\quad An ideal telescope would focus light from a distant source into a perfect diffraction-limited \ac{psf}. However, in practice, wavefronts are perturbed by the introduction of phase aberrations due to imperfections in optical surfaces, mirror alignment errors, and thermal distortions, demanding accurate wavefront sensing to determine a map of \acp{opd} over the entrance pupil. Once the \ac{opd} values are determined, the system can be corrected with either active or adaptive control, or the information can be used to inform post-processing of the data \citep{Hardy1998,2005Guyon,Tyson2015,Kenworthy2025}.

The \ac{jwst} relies on focal plane wavefront sensing, in which wavefront information is extracted from the same cameras used for science observations. Controlled defocus phase retrieval has been successful in determining pupil phases for intensity measurements under known perturbations \citep[e.g.][]{2012Acton,2022Acton}.  This has the advantages of minimising hardware requirements and avoiding non-common path errors between the science camera and a separate sensor. The same hardware can also be used for science programs \citep[e.g. defocused exoplanet transit photometry,][]{2023PASPSchlawin}.

The timescales involved for \ac{jwst} wavefront variations range from short-term variations caused by spacecraft slew and thermal settling, and longer-term variations due to structural relaxation and occasional impacts from micrometeoroids in the space environment around the observatory. Unlike many ground-based active telescopes, for JWST wavefront sensing and control cannot be done in parallel with science observations, but instead must be done as an engineering activity, taking typically an hour, thus adding an overhead which reduces time available for science. Therefore, wavefront control has to be considered as an ongoing activity where calibration overheads have to be balanced against scientific output. This requires robust methods that can perform well in regimes of limited signal-to-noise ratios and limited time sampling, while keeping the optical system within performance limits \citep{2012SPIEKnight,2014Feinberg}. 

\subsection{Wavefront sensing in JWST with NIRCam}
\label{sec:jwst_wfs}
\quad The JWST \ac{wfs} approach uses the science detector of the Near Infrared Camera \citep[NIRCam,][]{2023PASPRieke} as the primary sensory element. This strategy is made possible because of the excellent thermo-mechanical stability of the space environment, which has dominant aberration evolution timescales of hours to days rather than the $\sim$milliseconds encountered in ground-based contexts \citep{2022Acton,2023Lajoie}.

\ac{jwst}'s \ac{wfs} implements a hierarchical approach to bridge the dynamic range between post-deployment misalignment and the final diffraction limited performance. During commissioning after deployment, geometric and centroid-based techniques were used to locate and position the segment images. The large coarse piston errors (initially in excess of hundreds of microns) were then sensed with the Dispersed Hartmann Sensor, which measures relative segment piston based upon wavelength dependent interference patterns. Subsequently, global solutions were determined using least-squares reconstruction across all segments measured via phase retrieval with focus diversity via the secondary mirror. Multi-field aberrations were measured and corrected via a Hartmann-test-like process using unstacked segment centroids. \citep{2014Feinberg,2022Acton}. These steps were performed in commissioning and do not recur during science observations.

Following this, fine cophasing was performed using focus-diverse phase retrieval of NIRCam weak lens images. At this stage, the wavefront is effectively continuous which allows for reconstruction across the aperture to nanometre scale accuracy. Hybrid diversity algorithms permit this type of reconstruction by enabling adaptive feedback to enhance the robustness of convergence as well as diminishing dependence on explicit phase unwrapping \citep{2006SPIEDean}. The same fine alignment step has subsequently been repeated every few days to monitor and maintain the mirror alignments throughout science operations, keeping the total wavefront error typically at a root mean square (RMS) value of $\lesssim70 $ nm.  These algorithms deliver a sensing noise floor of about 5 to 7\,nm RMS and allow maintenance of alignment stability to a value better than ${\sim}$10\,nm over a multi-day interval \citep{2023PASPMcElwain}. Both JWST's total system wavefront error (WFE) RMS and its stability are significantly better than the observatory's design requirements (respectively, total WFE below 150\,nm RMS and stability better than 54\,nm RMS per 14\,days).

The \ac{wfs} system inverts focal plane intensity measurements to find the OPD across the segmented pupil. Phase-diversity measurements are employed to break the sign ambiguity in the recovered pupil phase. NIRCam uses weak lenses with known defocus offsets to deliver phase diversity. Image sets with known phase perturbations can then be collected \citep{2006SPIEDean,2023Lajoie} and phase retrieval solved using iterative algorithms. In operations, images are returned to Earth for processing, and converted to actuator commands that are sent back to the telescope only when the measured wavefront exceeds quantitative criteria indicating a correction is needed \citep{2022Acton, 2026Flagey}.

The \ac{jwst} \ac{wfs} operations software, the Wavefront Sensing System (WSS), is designed to monitor and correct the optical state of the observatory through a series of planned sensing and control actions \citep{2022Acton}. The algorithms used in the phase retrieval are intentionally conservative, with priority given to robustly retrieving controllable segment modes such as piston, tip and tilt, and to having high dynamic range to accommodate large initial wavefront errors in excess of a wavelength \citep{2006SPIEDean}, so that the system reliably converges towards an operating condition. This has contributed heavily to the success of the commissioning of the JWST and its long-term stability. This pipeline is, however, not publicly available as  open-source and therefore cannot be directly incorporated in an inverse framework to support science analysis. 

Phase retrieval algorithms fall into two broad categories, iterative transform methods (e.g. \citealp{Gerchberg1972APA}) and parametric optimisations. Both classes of algorithms have long heritage in astronomy, for instance being used in diagnosis and correction of the Hubble Space Telescope's spherical aberration \citep{1991Fienup, 1991Lyon}. The hybrid diversity algorithm used in the WSS is in particular an iterative transform method optimised for high dynamic range on segmented apertures.
During JWST commissioning, the mission operations team also used several additional phase retrieval packages employing a variety of architectures, including parametric optimisation via algorithmic differentiation \citep{2014Jurling, 2012Dean} and distributed computing on \acp{gpu}. In this context the WSS intentionally serves as a conservative baseline method. Parametric phase retrieval methods have continued to be used alongside the WSS for characterising JWST performance in science operations as well \citep{2024Telfer, 2026Telfer}. As each algorithm is subject to distinct biases and limitations, comparisons and cross-checks between independent tools are a proven approach for assessing systematics and validating overall physical correctness.

The approach taken in this work views wavefront sensing as a unified Bayesian inference problem. The method estimates both the pupil OPD and per-exposure nuisance parameters (e.g. stellar position, flux, background) simultaneously.  This formulation also allows the wavefront so recovered to be used in predictive modelling of PSFs across various observing configurations, simultaneously with other inference in any compatible simulations. The resulting framework is an open-source tool that supports application to \ac{jwst} but whose applicability can be expanded to other missions.

\camino (Computational Aberration Modelling \& Inference for NIRCam Observations) is an open-source Python package that we have developed to execute wavefront sensing and modelling of physical optics of JWST/NIRCam as part of this work. The package builds upon the ecosystem of the \dlux{} differentiable optics code \citep{2024arXiv240608704D} in the \texttt{JAX} language, the \equinox object-oriented \jax framework \citep{equinox}, and the \optax optimisation framework \citep{optax2020github}. Through these libraries we provide a complete differentiable framework for performing high dimensional optical inference. We have released the package on the PyPI (Python Package Index), in order to allow the community to install it via \texttt{pip install jwst-camino}, while also maintaining development versions at \href{https://github.com/rayshrish/camino}{github.com/rayshrish/camino} and documentation at \href{https://rayshrish.github.io/camino/}{rayshrish.github.io/camino/}. \camino is envisioned as a community resource and therefore, the community is invited to contribute by reporting bugs, requesting features or otherwise supporting its continued development.

In Section~\ref{sec:obsdata}, we discuss the observational data and preprocessing steps taken to prepare them for phase retrieval. In Section~\ref{sec:forward_model}, we describe the implementation of the forward model. In Section~\ref{sec:inverse_problem}, we discuss the optimisation and inference machinery, with a special focus on tilt events in Section~\ref{sec:tilt_events}. We present its outcomes in Section~\ref{sec:results}, and evaluate these in the context of the future of the mission and discuss extensions to this work in Section~\ref{sec:conclusions}.

\section{Observations and data}
\label{sec:obsdata}
\quad We comprehensively reanalyse the dataset of NIRCam wavefront sensing exposures taken as part of routine
observatory calibration from July 2022 to March 2026. The instrumental configuration and observational strategy are described in the following sections.

\begin{figure*}[t]
    \centering
    \includegraphics[scale=0.57]{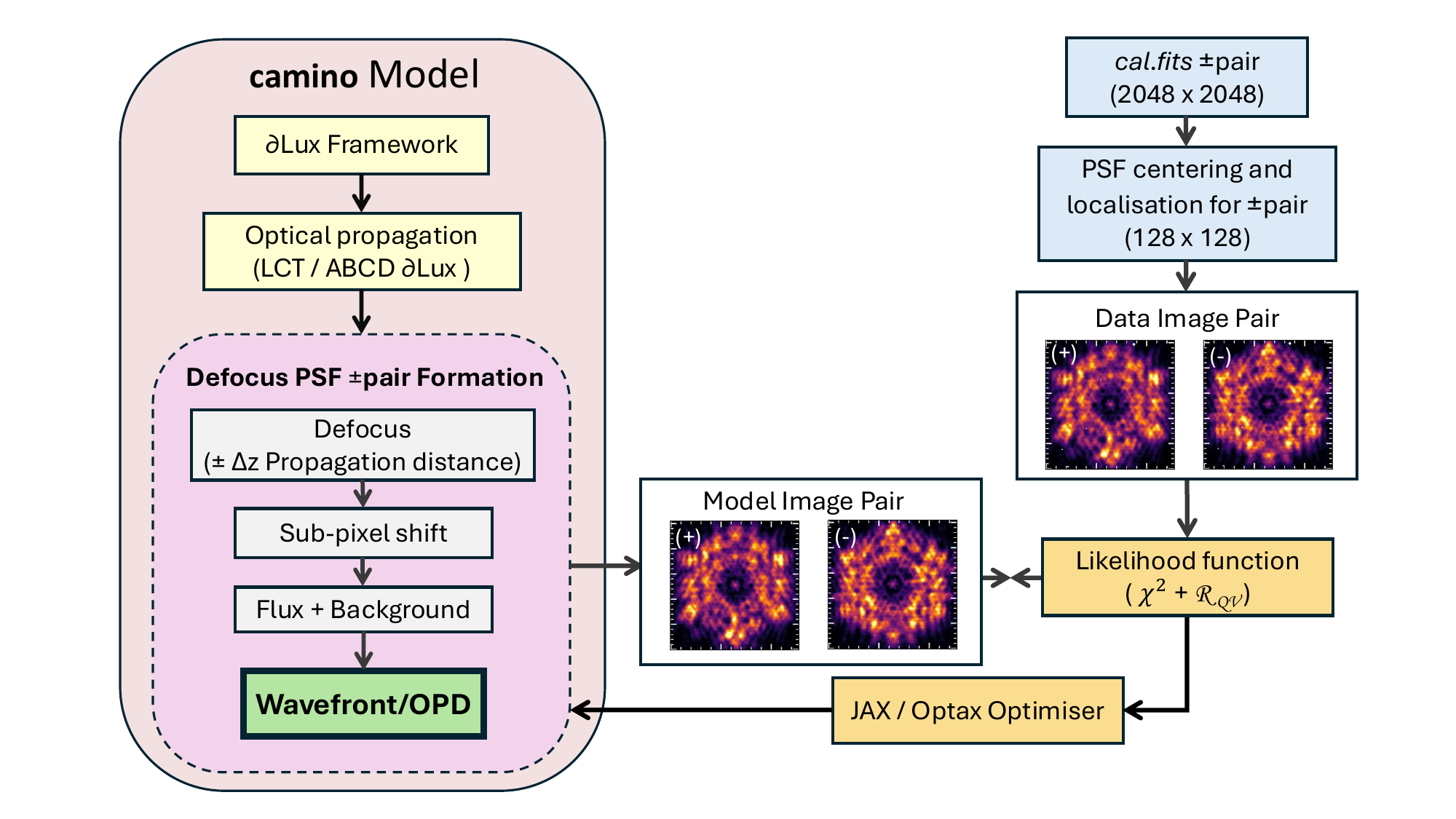}
    \caption{High level flow diagram showing the \camino framework. Each of the components are discussed in detail in Sections \ref{sec:forward_model} and \ref{sec:inverse_problem}.}
    \label{fig:camino_flow_chart}
\end{figure*}

\subsection{NIRCam weak-lens calibration observations}

\begin{figure*}[t]
    \centering
    \includegraphics[width=\linewidth]{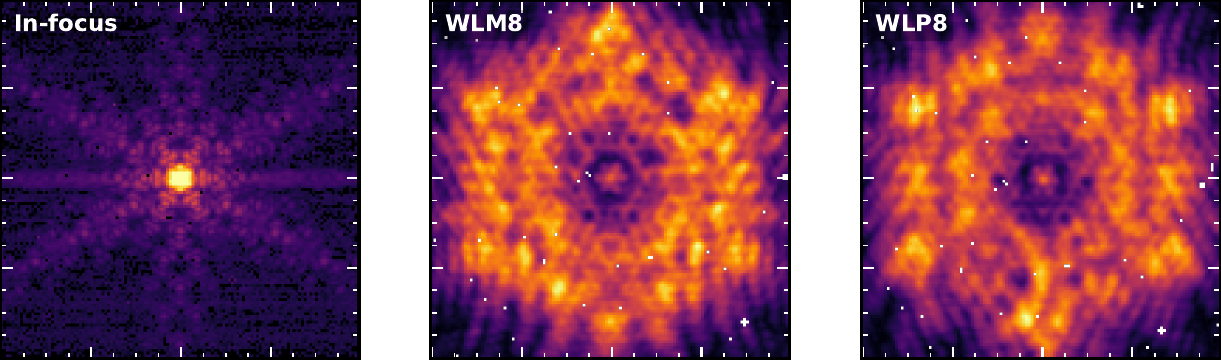}
    \caption{The in-focus NIRCam PSF is shown on the left panel from program 1160 (\ac{wfs} NIRCam Fine Phasing; target TYC 4212-1079-1, PI: M. Perrin). This was obtained with the F405N filter. The two defocussed NIRCam PSFs with WLM8 and WLP8 weak lenses are shown in the centre and the right panels respectively, from the program 2726 (Target: 2MASS-19512923-3023130; PI: M. Perrin). Both were obtained with the F212N filter. Each image has been placed on the same $128 \times 128$ pixel grid. Although these have varied instrumental configurations, it is clear that each defocussed PSF demonstrates the redistribution of intensity over a much greater spatial range than does each in-focus PSF.}
    \label{fig:psf_defocus}
\end{figure*}

\quad The observational data analysed  here use NIRCam’s F212N filter, which is a narrow band filter centred at $2.12\,\mu\mathrm{m}$. Passing a very narrow limited band, this filter largely mitigates chromatic effects in phase retrieval and was used to make pairs of images with weak lens elements WLP8 and WLM8, which produce similar amounts of defocus of nominally $\pm 8$ waves at a reference wavelength of $2.12\,\mu\mathrm{m}$ (see Figure~\ref{fig:psf_defocus}). The defocus produces a wide point spread function, the balanced defocus with opposite signs provides robustness against certain systematics, and the specific amount of defocus was chosen to optimise sensitivity on the spatial scale of JWST's segments and thus improves measurement accuracy \citep{2003JOSAADean,2016SPIEPerrin,2023PASPSchlawin}.

WLP8 and WLM8 together provide phase diversity and constraints on low-order aberration signs and magnitudes, and so improve the conditioning of the inverse solution. Because JWST’s wavefront evolves slowly, the WLP8/WLM8 image pairs can be treated as measurements of essentially the same underlying optical state over the timescale of the observation. 

\begin{figure}[h]
    \centering
    \includegraphics[width=\linewidth]{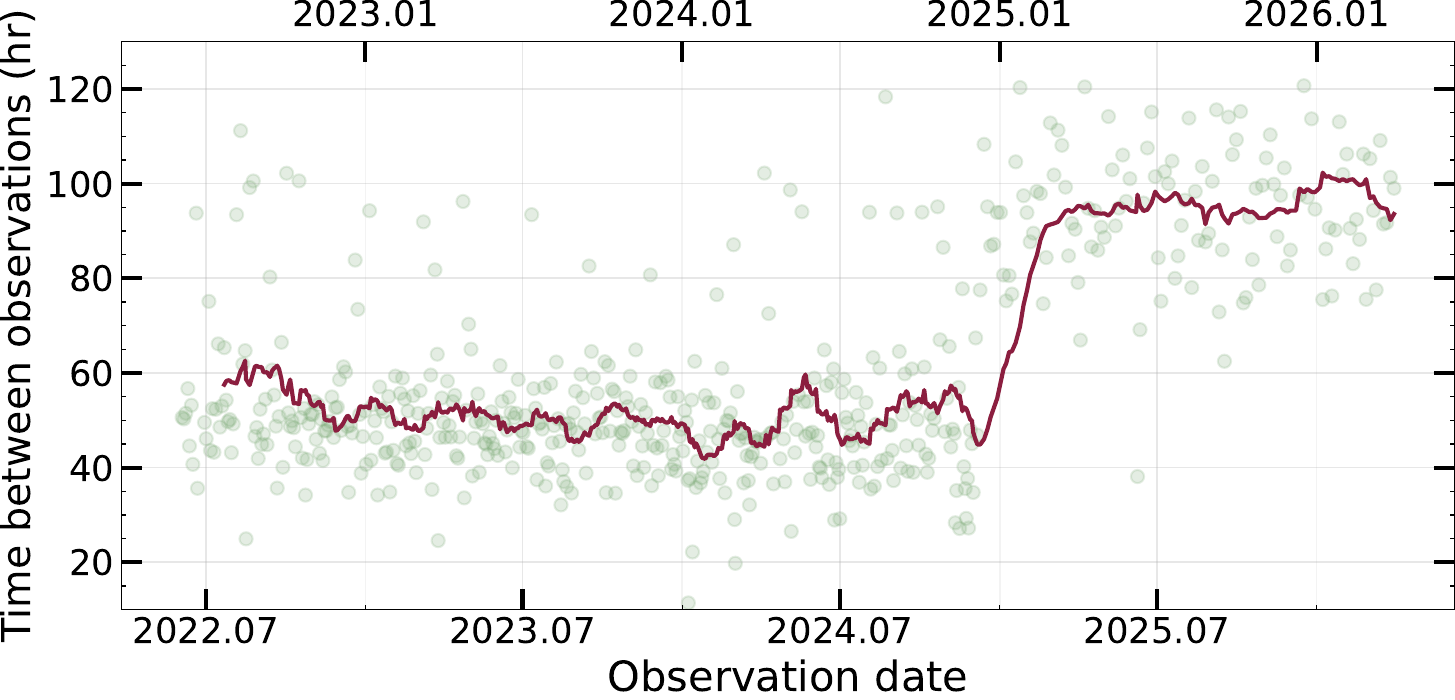}
    \caption{Time interval between consecutive JWST/NIRCam weak lens wavefront sensing observation pairs as a function of observation date shown with green circles, with a moving-average trend overlaid shown with a maroon line. The cadence remains ${\sim}40{-}60$ hours until late 2024, after which it increases to ${\sim}90{-}100$ hours.}
    \label{fig:wl_cadence}
\end{figure}

Routine wavefront sensing measurements have been taken in a series of calibration programs throughout JWST science operations. Various measurement parameters have been adjusted over time based on in-flight experience to optimise operations \citep{2026Flagey}. Initially, \ac{wfs} data were collected roughly every 2 days. Given the observatory's measured stability, the sensing cadence was reduced to every 4 days in December 2024 to reduce the sensing overhead time cost (see Figure \ref{fig:wl_cadence}). Another key change at that time was the relocation of the sensing field point from NIRCam detector A3 to detector A1, a change that mitigated occasional persistence effects in subsequent science observations. Initially observations were taken with 2 exposures per weak lens, dithered by 5 pixels to allow for bad pixel correction, but this proved unnecessary and was eventually ceased in order to reduce sensing time. The result of these changes is a time series of WFS measurements that are relatively but not entirely homogeneous, in particular with several intentional changes in sensing setup around 2024 December \citep{2026Flagey}.

\subsection{Data acquisition and selection}

\quad Each selected dataset associated with a wavefront measurement was processed into their calibrated detector level product (\texttt{cal.fits} file) produced by Stage~2 of the \ac{jwst} calibration pipeline \citep{bushouse_2025_17515973}. This ensured that the entire dataset contained the same corrections including flat fielding, photometry and astrometry so that they could be uniformly analysed as a representation of the intensity distribution. 

Each calibrated exposure was cropped to a fixed-size ($128\times128$) pixel sub-image centred on the defocused PSF using an iterative centroiding procedure. Where dithered observations were available, all WLP8 and WLM8 exposures were fitted simultaneously with a shared wavefront solution and per-exposure nuisance parameters.

\section{Forward Model}
\label{sec:forward_model}
\quad The focus-diverse observations as defined in Section~\ref{sec:obsdata}, are related to the pupil field through a known optical propagator. This is provided by a physically consistent forward model which maps the complex optical field through the system from the entrance pupil to the detector plane.

Wavefront sensing as outlined in Section~\ref{sec:jwst_wfs}, is considered as a unified inference problem. In this, the OPD at the pupil is treated as a common latent variable for multiple observations. The forward model provides a generative description of the instrument allowing prediction of the expected images given a particular wavefront realisation while taking into account both instrumental and observational effects.

This section defines the forward operator employed throughout. First, we describe the relationship between the complex field at the pupil and the OPD. Next, we explain how this field propagates to form an image in the detector plane. Following that, definitions of the models describing the formation of an image, inversely solving the problem using regularised optimisation methods and related techniques are provided. For clarity, we present a flow diagram of the method in Figure \ref{fig:camino_flow_chart}.

\subsection{Pupil-plane field representation}

\quad With the assumption that scalar diffraction applies, the electric field can be represented as
\begin{equation}
U(x, y) = A(x, y)\,\exp{\left[i\,\varphi(x, y)\right]},
\end{equation}
where $(x, y)$ are the pupil-plane coordinate pairs, $A(x, y)$ defines the amplitude transmission function, and $\varphi(x, y)$ defines the phase function.

The phase function is defined with respect to the OPD as  
$\varphi(x, y) = 2\pi/{\lambda}\cdot\mathrm{OPD}(x, y),$
where $\lambda$ represents the effective wavelength of observation. Here, the OPD provides a physically meaningful way to describe the wavefront error in relation to an ideal reference surface. 

The amplitude component represents the physical and transmissive characteristics of the telescope aperture including the geometry of the segmented primary mirror, supporting spider tripod and the central obstruction. The transmission $A(x,y)$, is assumed to be unity across the pupil's support. Additional residual throughput variation which may occur due to optical errors or misalignment is also accommodated. Although the amplitude is treated as uniform over the pupil, we anticipate that in more accurate treatments it will vary at spatial scales comparable to those associated with the pupil geometry and the overall optical design of the telescope. The specific amplitude transmission map $A(x,y)$ used in this work is based on the model from \texttt{stpsf} \citep{2014SPIEPerrin}, with additional empirical refinements estimated from in-flight NIRCam pupil imaging measurements. These refinements better reproduce the geometry of the spider of the telescope. This transmission map may have uncertainties (e.g. related to inter-segment distances and edge effects), which can be improved by optimising for the illumination of the pupil in future work (see Section \ref{sec:model_improvements}).

The phase term $\varphi(x,y)$ contains the optical aberration terms and represents the primary parameter to be inferred. For segmented telescopes, like \ac{jwst}, there exist intra-segment smooth wavefront structures and inter-segment discontinuous wavefront features caused primarily by piston, tip and tilt misalignments. Therefore, our choice of phase representation does not impose a particular modal base on the wavefront representation but instead allows the preservation of the complete spatial structure of the wavefront for direct recovery of both lower and intermediate order aberrations from the measured data.

\subsection{Optical propagation model and defocus}
\label{sec:optical_propagation}

\quad We make the simplifying assumption that JWST's three mirror anastigmat optical design can be modelled as a single transform from a pupil plane to an image plane.

To relate pupil-plane fields to focal-plane intensity, the scalar diffraction method is used to propagate the complex optical field through the system. Propagation from pupil to focal plane is modelled via a Collins diffraction integral, which uses a \ac{lct}, that relies on an ABCD ray transfer matrix \citep{1970JOSACollins,2005ifoGoodman}.

For a thin lens followed by free-space propagation, the matrices for each component of the system are
\begin{equation}
M =
M_{\mathrm{free}}(L)\cdot M_{\mathrm{lens}}(f)
=
\begin{pmatrix}
1-\frac{L}{f} & L\\
-\frac{1}{f} & 1
\end{pmatrix}
\end{equation}
where $f$ is the focal length and $L$ is the propagation distance. We note that JWST's telescope optics are implemented with reflective mirrors, not a thin lens, but the same equation applies.

In order to apply a numerical Fourier Transform to a differentiable matrix, the Collins integral is computed. As such, the input and output sampling are independently specified.

Defocus is related by one effective propagation distance offset,  $\Delta z$. This represents a relative shift of the focal plane from the detector, imitating the defocus produced by the weak-lens optics. Under nominal focusing conditions, $L{=}f$. A defocus is then just
\begin{equation}
L = f + \Delta z.
\end{equation}

\noindent and the system matrix under defocused conditions is therefore
\begin{equation}
M(\Delta z)
=
M_{\mathrm{free}}(f+\Delta z)\cdot M_{\mathrm{lens}}(f)
=
\begin{pmatrix}
-\frac{\Delta  z}{f} & f+\Delta z\\
-\frac{1}{f} & 1
\end{pmatrix}.
\end{equation}

The implementation of the optical model is carried out within the \jax framework using the \dlux and \textsc{Zodiax} library \citep{2023JATISDesdoigts,2024arXiv240608704D}, built on \textsc{Equinox} \citep{equinox}, for end to end differentiable optical modelling. This framework has also been successfully applied in recent analyses of \ac{jwst}'s Near Infrared Imager and Slitless Spectrograph (NIRISS) aperture masking interferometry (AMI) observations \citep{2025Desdoigts, 2025Charles}.

\subsection{Detector and Wavefront Sampling}
\label{sec:detector_sampling}

\quad The focal plane field propagated through the optical system is discretised onto an oversampled detector grid (factor of $s=4$), then downsampled to the native resolution of the detector. The model PSF is normalised so that it has unit flux, while each exposure is represented as
\begin{equation}
I_{\mathrm{obs}}(x,y)=f\cdot P(x,y)+b
\end{equation}
where $f$ is the source flux, and $b$ is the uniform background illumination. Since both $f$ and $b$ are linear terms, we can solve for them analytically at each model evaluation using weighted least squares, with the per-pixel uncertainties from the calibrated data product. 

Polychromatic calculations are performed by summing over $N$ monochromatic propagations. Specifically the narrow but nonzero width F212N filter is modelled by summing over 5 wavelengths.

\section{Inverse problem and regularisation}
\label{sec:inverse_problem}

\quad The forward model maps a set of parameters $\Theta$ to detector plane images
\begin{equation}
I_{\mathrm{model}}^{(k)} = \mathcal{F}\left(\Theta; k\right),
\end{equation}
where $k$ represents individual exposures. $\Theta$ consists of a shared OPD and exposure specific quantities (source position and defocus). 
The inverse problem is stated as the minimisation of a regularised loss function
\begin{equation}
\mathcal{J}(\Theta) = \sum_{k} -\mathcal{L}_{k}(\Theta) + \mathcal{R}(\Theta),
\end{equation}
\noindent where $\mathcal{L}_{k}$ is the log-likelihood, and $\mathcal{R}$ is a composite regularisation functional.
We assume a pipeline-stated Gaussian distribution of errors at each pixel location $\sigma(x,y)$ and use the corresponding Gaussian log-likelihood. 
Invalid pixels are masked out using a binary mask generated from the calibrated science image. A pixel is considered invalid if it contains NaN. We analytically find and marginalise over flux and background values \citep{Luger2017} to ensure that the likelihood depends only upon the remaining latent variables. 

\subsection{OPD parameterisation and regularisation}
\label{sec:regularisation}

\quad The OPD is modelled as a scalar field $\phi(x, y)$. It is mean-subtracted across the pupil support to eliminate the piston degeneracy.

Smoothness of the OPD is ensured using a segment-wise quadratic variation (QV) penalty, proportional to
\begin{equation}
\mathcal{R}_{\mathrm{OPD}} = \lambda_{\mathrm{OPD}} \sum_{s} \sum_{(i,j)\in s}
\left[
(\Delta_x \phi)^2 + (\Delta_y \phi)^2
\right].
\end{equation}
\noindent where $\Delta_x$ and $\Delta_y$ refer to first finite differences in each dimension. This is computed separately over the interior of each segment, ignoring jumps between segments.  Note that $\lambda_{\mathrm{OPD}}$ here is a regularisation parameter, and is unrelated to the wavelength $\lambda$.

\subsection{L-curve selection of the regularisation parameter}
\begin{figure*}[h]
    \centering
    \includegraphics[scale=0.65]{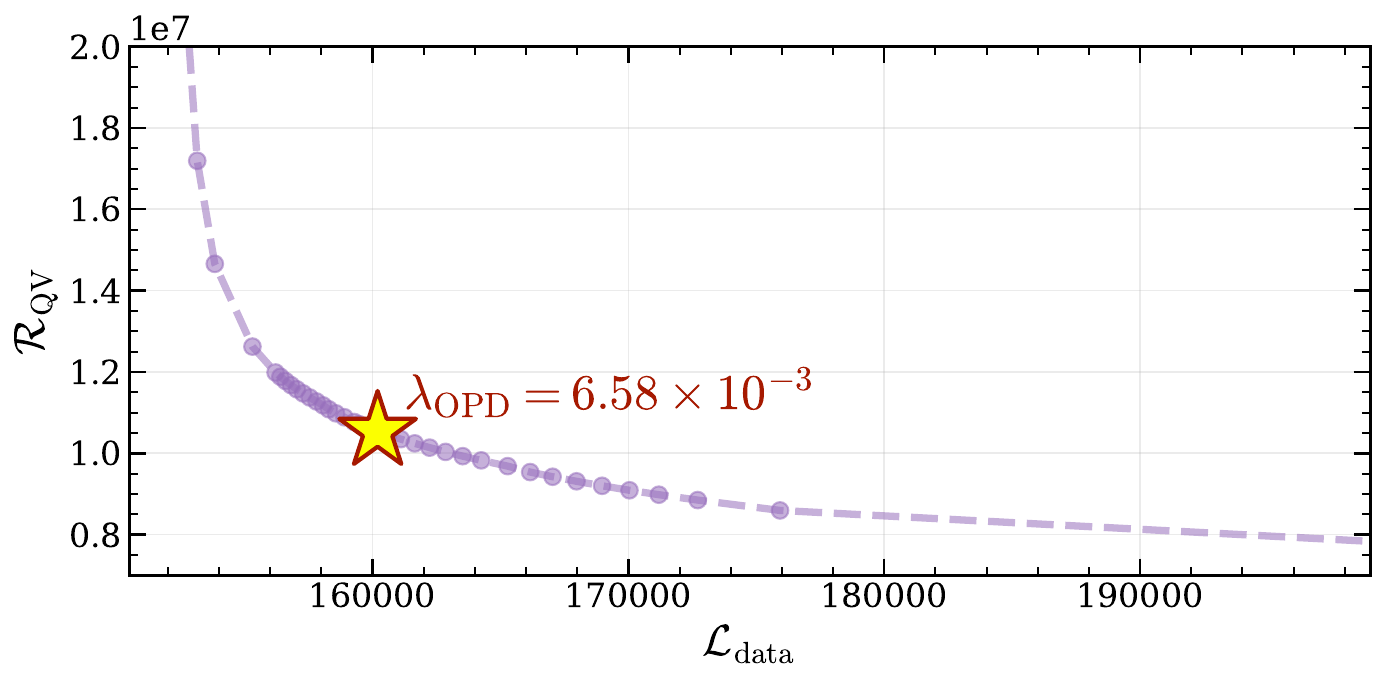}
    \caption{L-curve for the segment-wise quadratic variation (QV) regularisation, showing the trade-off between data misfit ($\mathcal{L}_{\text{data}}$) and regularisation penalty ($\mathcal{R}_{\text{QV}}$). A coarse sweep (shown in light purple) followed by refined sampling (shown in dark purple) around the transition region is shown. The chosen value, $\lambda_{\mathrm{OPD}} = 6.58 \times 10^{-3}$ (shown with a yellow star) lies at the knee, the point at which the strength of the smoothness and the quality of the fit is balanced.}
    \label{fig:opd_l_curve}
\end{figure*}

\quad The value of $\lambda_{\mathrm{OPD}}$ controlling the smoothness of the QV penalty was determined by an L-curve analysis \citep{Hansen2001}. For a particular choice of the regularisation strength $\lambda_{\mathrm{OPD}}$, we solve the optimisation problem and evaluate $\mathcal{R}_{\text{QV}}$ and the negative log-likelihood $\mathcal{L}_{\text{data}}$ (see Figure~\ref{fig:opd_l_curve}). The L-curve is created in two logarithmic sweeps:
\begin{itemize}
    \item a coarse sweep over $\lambda_{\mathrm{OPD}} \in [10^{-4}, 10^{0}]$,
    \item a refined sweep over $\lambda_{\mathrm{OPD}} \in [3\times10^{-3}, 1.5\times10^{-2}]$,
\end{itemize}
with all other parameters are kept constant. Fitting cubic smoothing splines to the refined grid data and maximising its geometric curvature, the optimal regularisation strength was calculated to be $\lambda_{\mathrm{OPD}} = 6.58\times10^{-3}$ (see Figure \ref{fig:opd_l_curve}). This value indicates the best trade off between data fidelity and regularisation \citep{Hansen1992_lcurve}. Additionally, by reducing the effect of low confidence, rapidly varying structure within the OPD solution space, the QV regulariser limits the number of degrees of freedom available to the model. Therefore it limits an excessively flexible high-dimensional parameterisation which would be conducive to overfitting \citep{2021Hogg}. 

\subsection{Optimisation strategy and implementation}

\quad The objective function defined above is minimised using the Adam optimiser in \jax/\optax, with gradients calculated with automatic differentiation. To reduce the coupling between geometric and wavefront parameters, we employ two stages of optimisation. In the first stage, only the exposure-specific parameters (position and defocus values), are optimised whilst the OPD perturbations are held fixed. This stage results in the alignment of the model with the observed data by mitigating large scale mismatches in image position and focus. In the second stage, the exposure-specific parameters are held fixed and the OPD map is optimised.

During the two stage optimisation, convergence is assessed using the relative improvement ${\left| \mathcal{J}_{t} - \mathcal{J}_{t-\Delta} \right|}/{\mathcal{J}_{t}}$, where $t$ is the iteration index and $\Delta$ is the window size, set to 50. 
Optimisation terminates once the relative improvement has decreased below a specified threshold of $10^{-6}$ or if there has been a maximum of 5000 iterations. These criteria were determined empirically and resulted in consistently stable convergence behaviour across all analysed data sets. All computations were vectorised and used just-in-time compilation provided by \jax to improve efficiency.

\section{Mirror Segment Tilt Events}
\label{sec:tilt_events}
\quad 

There exists a subset of epochs where there is an abrupt change in the optical state related to movement of a segment of the \ac{jwst} primary mirror \citep[e.g.,][]{2022Acton, 2024Perrin, 2024Telfer}. These ``mirror segment tilt events''  create significant segment-scale disturbances rather than the usual gradual evolution of the wavefront. These are not as efficiently recovered by pixel-basis models as geometric models explicitly including the segment tilt, which we have incorporated into \camino to ensure an \ac{opd} solution can be obtained at any epoch. The following sections explain this approach in detail.

\subsection{Event Detection}

\quad Candidate mirror segment tilt events which represent a significant change in optical state are identified using preliminary OPD solutions of the entire data set. The OPD for each observation is compared with the OPD of the previous observation. This comparison provides an independent evaluation of the wavefront error at each primary mirror segment. A measure of the size of the inter-epoch change is determined by calculating the RMS value of the change at each segment. Epochs where one or more segments have experienced an exceptionally large RMS value change are considered candidate mirror segment tilt events. A threshold of $200\,\mathrm{nm}$ was selected based on the empirical analysis of the distribution of RMS values over all epochs of the data base. A distinct class of large amplitude wavefront disturbances is found which are separate from the nominal background evolution of the telescope wavefront (in practice, tilt events are known to occur on a continuum, including events below this threshold size; see \citealp{2024Telfer}).

Tilt events usually occur abruptly and last over several consecutive observations (${\sim}2$ epochs) before being corrected. However, the precise onset and duration of these events are limited by the observational cadence, and shorter-lived variations occurring between observations may remain undetected. Shorter-cadence measurements have revealed wavefront variations on much shorter timescales, including rapid segment tilt events \citep[as discussed in Section 4.5.4 of][with some events lasting for ${<}10\,\mathrm{s}$]{2023Rigby}. Since we are retrieving the wavefront at given epochs, these type of variations do no affect our reductions.

Candidate detections are visually inspected and compiled into a final event catalogue to ensure that the identification of each physical event is consistent among all epochs of the observation. This inspection includes examining the recovered OPD map and its corresponding PSFs to verify if a segment scale signature is present. Each event in the catalogue contains information about when it occurred, which segments were involved, and which epoch represented the last non-event observation. This catalogue served as the foundation for the development of a specialised retrieval procedure.

\subsection{Event Retrieval}

\quad Optimising the tilt event models for retrieving the dominant displacement of each segment was achieved using a multi stage process and is explained in the subsections below.

\begin{figure*}[h!]
    \centering
    \includegraphics[width=\linewidth]{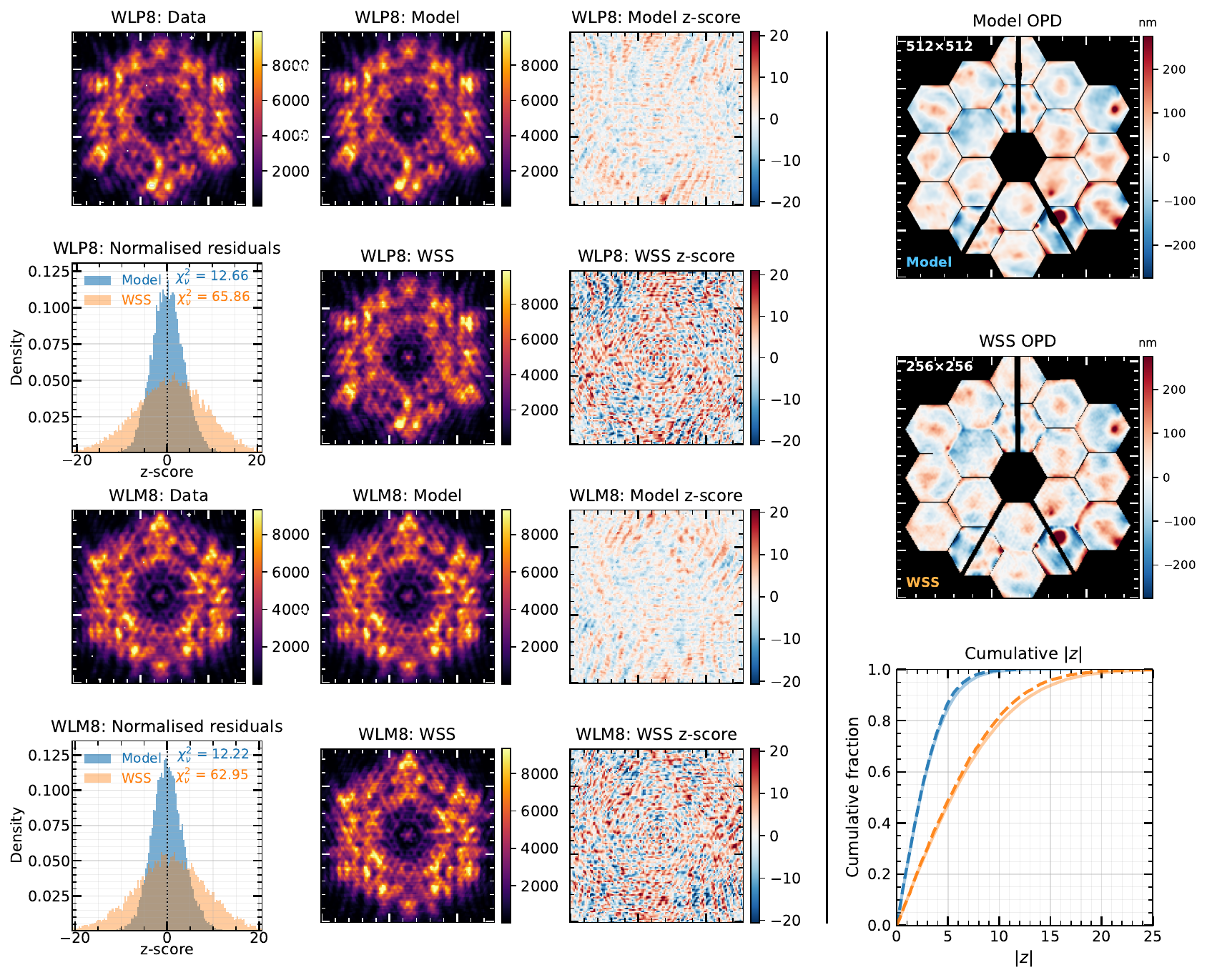}
    \caption{
    Comparison between the forward model presented in this work and the official WSS solution for NIRCam defocused images from GO~7464 (PI: T.~Beck). For each pupil (top two rows: WLP8; bottom two rows: WLM8), the left three columns show the observed data, the best-fitting forward model, and the corresponding normalised residuals expressed as $z$-scores, $z=(D-M)/\sigma$. The second row for each pupil contains the distribution of residuals together with the model image and residuals obtained using the contemporaneous WSS solution. Within each pupil, the data and model images share a common intensity scale, while the residual maps are shown on a common symmetric colour scale to facilitate direct visual comparison. Our forward model provides a substantially improved description of the observed PSFs, yielding narrower residual distributions and significantly lower reduced chi-squared values ($\chi_\nu^2\approx12${-}$13$) than the WSS solution ($\chi_\nu^2\approx60${-}$65$). The right column compares the corresponding wavefront solutions and their residual statistics. The upper panels show the retrieved OPD map from the joint WLP8/WLM8 fit and the contemporaneous WSS OPD estimate, both displayed on their native grids (indicated in the panels) with identical colour scales. The lower panel shows the cumulative distribution functions of the absolute normalised residuals, $|z|=|(D-M)/\sigma|$, aggregated over both defocus positions. Solid curves include all valid pixels, while dashed curves are restricted to the brightest 15\% of pixels. Our retrieved OPD solution (blue) exhibits substantially smaller residuals than the WSS solution (orange), including in the highest signal-to-noise regions, demonstrating a significantly improved description of the observed PSFs.
    }
    \label{fig:model_comparison}
\end{figure*}

\subsubsection{Grid search initialisation}
\quad The primary goal of this stage is to produce a robust initial estimate of the dominant segment displacement using a simple grid search. By explicitly examining the low dimensional parameter space of tip-tilt, this optimisation process is less prone to local minima compared to gradient-based methods starting from zero displacement.

\subsubsection{Joint segment plane optimisation}

\quad After the grid search initialisation, the segment piston, tip and tilt coefficients are then jointly refined using a BFGS \citep{Broyden1970, Fletcher1970, Goldfarb1970, Shanno1970} optimiser provided through \textsc{Scipy}'s \texttt{optimize.minimize} function. 

After convergence of the segment coefficients, the source positions are re-optimised while holding the recovered tip-tilt solution constant. Since large segment displacements can induce small shifts of apparent image centroids, a second position fit ensures reduced coupling between geometric registration errors and wavefront errors. Finally, the refined source positions are held constant whilst the residual wavefront optimisation is executed in the final stage as explained below. 

\subsubsection{Residual pixel-basis refinement}
\quad Once the segmental tip-tilt solution was determined, the baseline OPD and recovered tip-tilt coefficients were fixed. A residual pixel based correction was introduced and optimisation continued with a three-component wavefront model adding together the most recent non-tilt event wavefront state, a linear geometric model of the segment displacement due to the tilt event, and a pixel-basis map as usual for residual wavefront structure not represented by the segment model. 

At present in the \camino implementation, the retrieval framework is employed at all epochs. As a result, all segment tip-tilt parameters are always included in an optimisation, regardless of whether or not a mirror segment tilt event was recorded. As a result, for nominal observing epochs, the recovered tip-tilt amplitude values are generally negligible and are dominated by the standard pixel-based representation of the wavefront. The OPD maps obtained through this process were then used to construct a chronological record of the telescope's wavefront history over the entire observing sequence. To obtain the pupil RMS wavefront error for each epoch, the RMS value of the pupil (as defined above) is calculated directly from the final retrieved OPD map.

\begin{figure*}[h]
    \centering
    \includegraphics[scale=0.55]{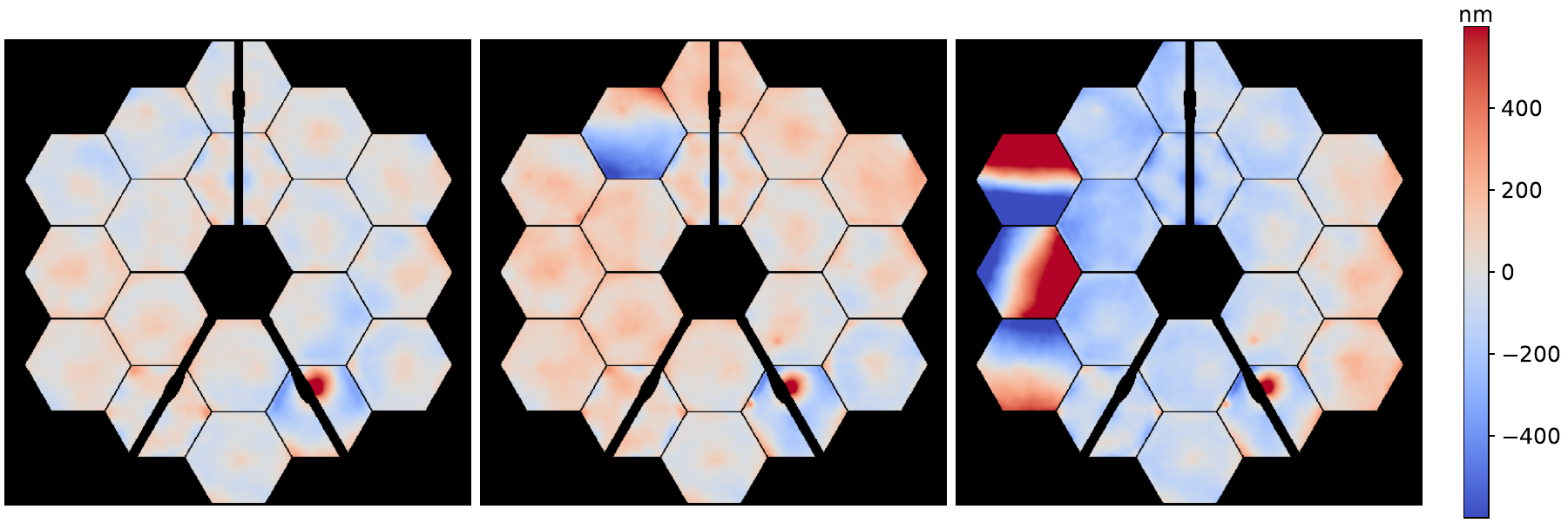}
    \caption{The left panel shows an OPD map without any tilt events. The middle panel shows a wavefront epoch with a tilt event in one segment (C6) and the right panel shows an epoch with a tilt event in multiple segments (B6, C5, and B5). The figure shows that tilt events cause significant RMS changes when compared to the nominal wavefront evolution.}
    \label{fig:tilt_events}
\end{figure*}

\begin{figure*}[h!]
    \centering
    \includegraphics[width=\linewidth]{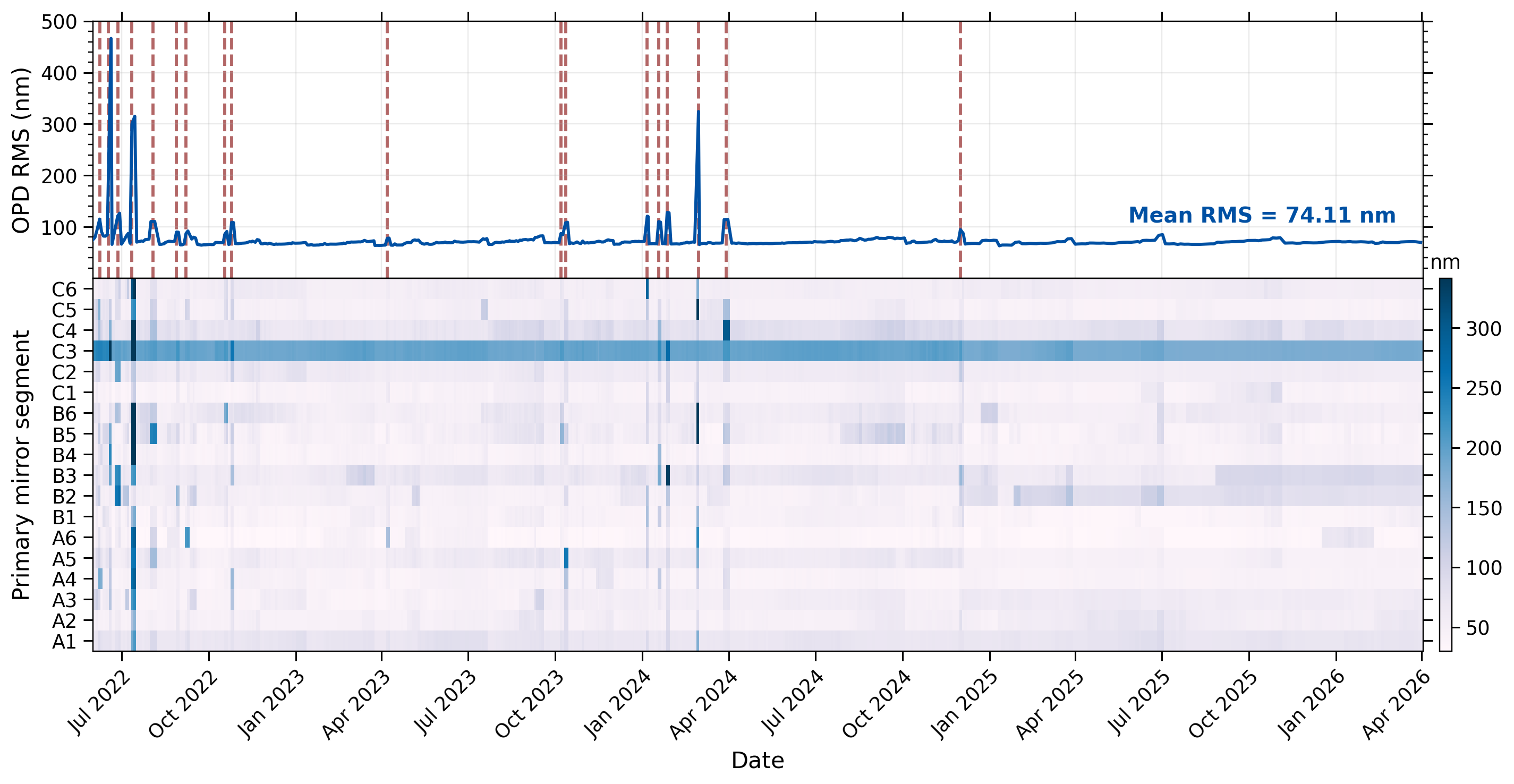}
    \caption{
    \textbf{Top panel:} Temporal evolution of RMS wavefront error (computed from the retrieved OPD maps). Dashed vertical lines show the identified mirror segment tilt events above our chosen threshold of 200 nm RMS. Over all of the observed data set, the mean RMS wavefront error is $74.11\,\mathrm{nm}$. \textbf{Bottom panel:} Temporal progression of RMS wavefront error separately calculated for each of the JWST primary mirror segments (named using the official labels). Segments correspond to rows, while the columns represent observation epochs. In this colourmap, darker colours indicate higher RMS wavefront error values at a given epoch. Many of the short-duration wavefront excursions coincide with the tilt events in the top panel. Due to the permanent wavefront distortion caused by the micrometeoroid event documented in May 2022, segment C3 shows consistently high RMS levels throughout the measurement period. Except for those events described above, the majority of segments have exhibited relative stability over their respective baselines. In December 2024, a change in sensing field point within NIRCam is visible as a systematic apparent change in several segments (actually a change in NIRCam internal WFE at the sensing field point); by coincidence there was also an unrelated tilt event on 1 December 2024. 
    }
    \label{fig:opd_rms}
\end{figure*}

\section{Results and Discussion}
\label{sec:results}

\quad The method presented in this work allows a direct estimation of a JWST OPD from a pair of weak lens PSFs in order to measure the telescope wavefront. We compare the estimated wavefront with the standard WSS products retrieved via \texttt{stpsf}, including the OPD solution and the model PSFs obtained with the WSS software in both WLP8 and WLM8 lenses. Furthermore, we analyse the progression of the wavefront over time and identify discrete events, in particular tilt events and micrometeoroid impacts.

\subsection{Comparison with Operational WSS Products}
\quad The WSS data products offer a common basis for comparing our forward model against WSS calibrated data. For this purpose, residuals were calculated relative to the calibrated data, using per-pixel uncertainty values, and expressed as normalised residuals or $z$-scores, $z = (\text{data} - \text{model}) /  \text{uncertainty}$. The residual maps corresponding to both defocus positions are provided in Figure\,\ref{fig:model_comparison}, along with their $z$-score distributions.

The residual maps from \camino are largely featureless. Conversely, the residuals generated from the use of the WSS products show clear spatial correlations. These seem most prevalent in regions of high intensity, thereby potentially implying that there exists some systematic mismatch between the predicted PSF and the measurement. These characteristics can be seen in quantitative detail through the distributions of $z$-scores. The residuals obtained using our model are clustered tightly about zero. However, the $z$-scores from WSS products have a broader distribution which can be attributed to an apparent ringing effect in the residuals, indicating persistent model discrepancies.

These differences are captured in the reduced $\chi^2$ statistics. For both WLP8 and WLM8, our retrieved model achieves values of $\chi^2_\nu \simeq 12$-13, while the corresponding values obtained using the WSS products are larger by factors of $\sim$5. Although both the models formally underfit the data relative to the ideal $\chi^2_\nu \approx 1$ expectation, the relative improvement is substantial and consistent across both defocus positions. Moreover, it can be noted that this improvement is not limited to low-signal regions, and persists in the brightest pixels.

The cumulative distribution functions (CDFs) of $|z|$, which are given in the right panel of Figure\,\ref{fig:model_comparison}, show the behaviour described above. It is clearly seen that the CDF of our retrieved model increases more steeply than the one for the WSS-based model, indicating that there are fewer pixels with large residuals. Additionally, the difference is more pronounced in the top 15\% brightest pixels. Here, we observe that the WSS-based model has many more large residuals than our retrieved model. Therefore, it is reasonable to conclude that the increase in accuracy in our retrieved model does not result from considering noisy regions, but rather from providing a better description of the actual wavefront.

We also compare the retrieved OPD map with the WSS solution obtained via \texttt{stpsf}. On a qualitative level, both maps appear to describe the same major features as far as segmental variations and general shapes are concerned. However, our retrieved OPD appears smoother than the one retrieved from WSS. As explained earlier, this smoothness arises from the application of the quadratic variation regularisation during inference. Despite this, the dominant wavefront structures and their corresponding segmental variations have been reasonably well reproduced.

\subsection{Wavefront Evolution and Mirror Segment Tilt Events}

\quad Using the retrieval framework for all the
wavefront sensing epochs produces a continuum of the telescope wavefront over time, starting from the commissioning phase through early 2026. Each epoch was used to calculate the RMS wavefront error using the retrieved OPD map, resulting in a time series shown in the top panel of  Figure~\ref{fig:opd_rms}.

The overall behaviour is characterised by an extremely stable wavefront punctuated by a few discrete excursions. The average RMS wavefront error over the total observational time frame is $74.11\,\mathrm{nm}$.  Note that this is the inferred total observatory WFE, including NIRCam instrument WFE as well as the telescope WFE. We do not attempt to subtract the NIRCam instrumental component from the overall wavefront in this version of \texttt{camino}, though this is done by both the WSS and \texttt{stpsf}. Most of the epochs are clustered closely around that value. Overlaid on that stable baseline are some high RMS wavefront errors, the most notable being those during the commissioning phase in the middle of 2022 and at the beginning of 2024.

Some examples of retrieved mirror segment tilt events are presented in Figure \ref{fig:tilt_events} and all such instances have been indicated with the maroon dashed lines on the top panel of Figure~\ref{fig:opd_rms}. These correspond to the largest excursions in RMS history. In each case, the wavefront errors increase rapidly because of segment displacement, and then return to their normal state following a correction.

A key feature of the RMS history is the absence of any significant long-term deterioration. The consistent optical performance demonstrated by the telescope reflects the long term stability of \ac{jwst}'s wavefront as well as the effectiveness of the sensing and control system.

\subsection{Segment-level stability and micrometeoroid impacts}

\begin{figure*}[h]
    \centering
    \includegraphics[scale=0.42]{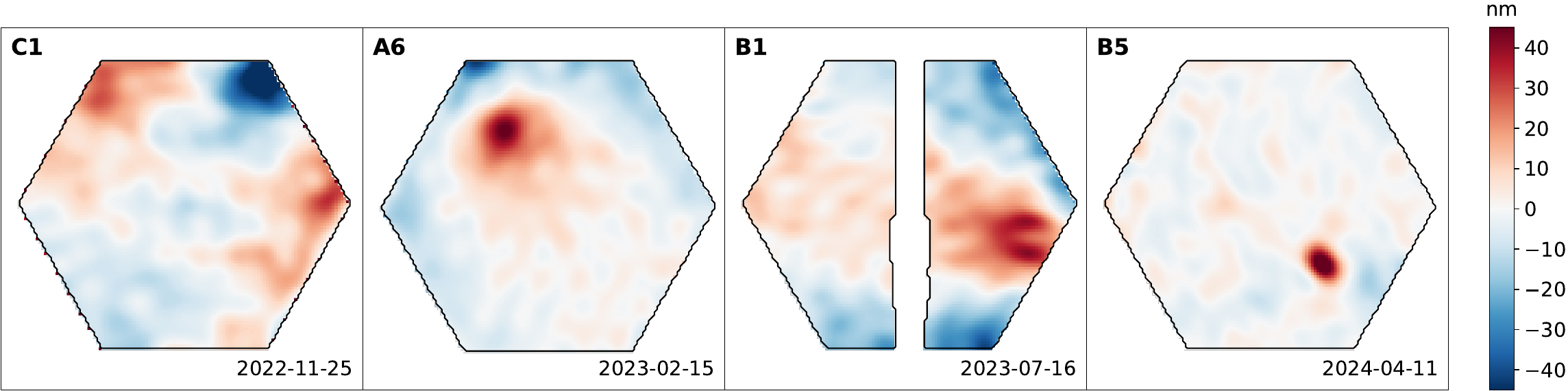}
    \caption{Example shapes of micrometeoroid impact effects. From left to right: a diffused negative impact, a diffused positive impact, a diffused positive impact with a low amplitude core, and a highly localised sharp impact. The corresponding date and segment label is also mentioned for each case.}
    \label{fig:impact_features}
\end{figure*}

\begin{figure*}[h]
    \centering
    \includegraphics[scale=0.5]{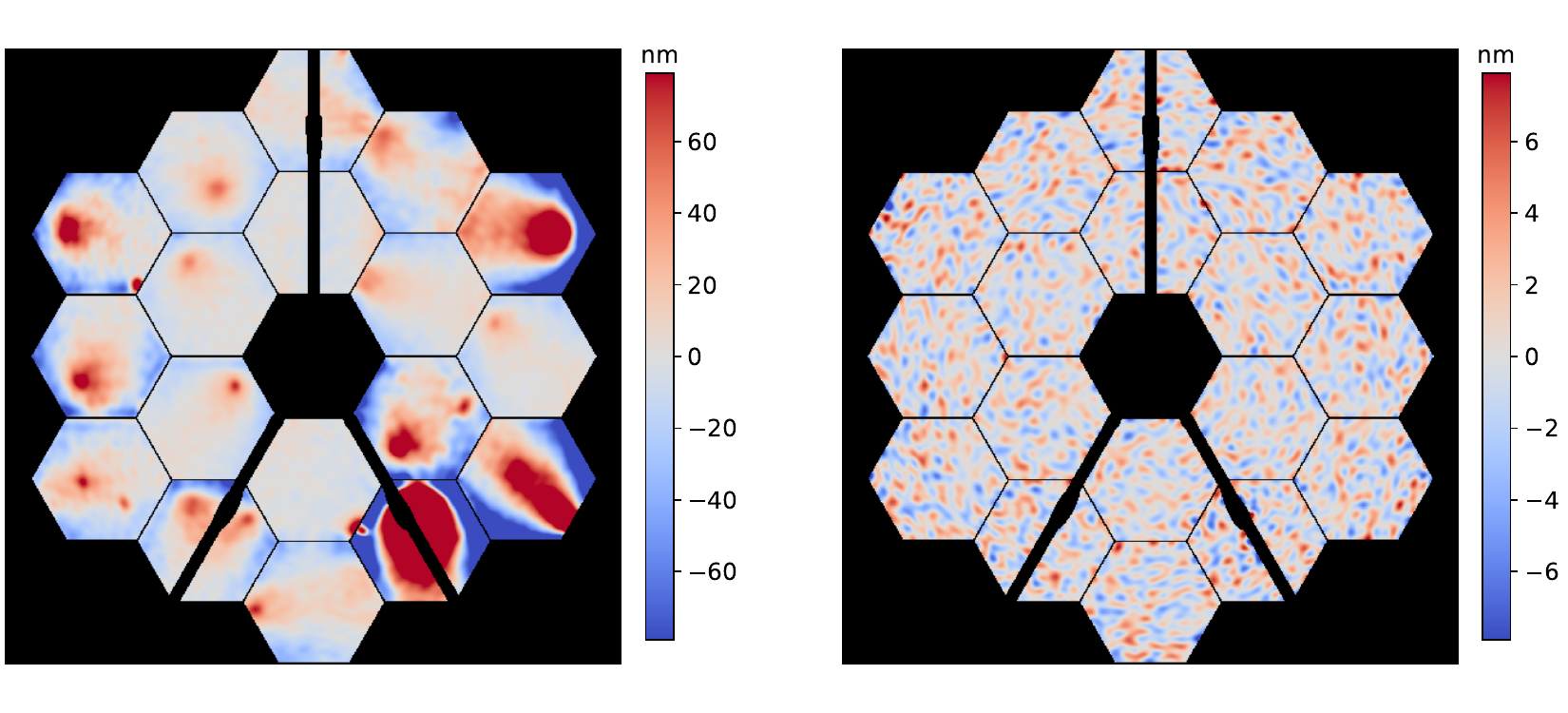}
    \caption{Left: Summary of all the micrometeoroid features made by adding all the impacts for each affected segment. Right: The difference between two typical wavefront retrievals in consecutive epochs in the absence of any tilt events or micrometeoroid impacts. This differential OPD residual represents a combination of true temporal wavefront evolution and fitting noise between consecutive epochs.}
    \label{fig:meteoroid_and_stability}
\end{figure*}

\quad The stability and temporal evolution of JWST's wavefront have previously been described by the mission operations team \citep{2023PASPMcElwain, 2024Telfer}, including the measured effects of micrometeoroid impacts on segments \citep{2026Telfer}. We here independently reconstruct some of these results as a demonstration of the efficacy of \camino. 

Although the global RMS wavefront history offers an overall summary of the performance of the telescope's optical state, it does not reveal how individual  mirror segments influence the observed evolution. Therefore we calculate the RMS wavefront error individually for each mirror segment at each epoch. This segment-level history has been presented as the bottom graph of Figure~\ref{fig:opd_rms}.

Most segments behave in broadly the same way. RMS wavefront error remains relatively constant over the baseline, but there are some occasional large transients which occur due to mirror segment tilts. These typically affect only a few segments at most, and the RMS returns to its original level once corrective control has been applied. 

In the bottom panel of Figure~\ref{fig:opd_rms}, a particularly significant constant elevated RMS can be seen in segment C3. This is consistent with the well-documented micrometeoroid impact that occurred in May 2022, causing a permanent localised deformation. Also, unlike the transient tilt events, the signature of this impact can be seen to persist throughout the entire observation period. Smaller but significant impacts on segments B2 and B3 can be seen in early and late 2025 respectively.

Beyond the persistent C3 feature, segment level RMS histories show that there has been no major decline in individual segment quality over time. This finding is consistent with previous works \citep{2023Lajoie, 2024Telfer} examining the global RMS history. In addition to this, it is possible to see smaller segment level features by comparing the difference between the phase retrievals from two subsequent epochs. By taking the difference between two such OPD maps and removing the piston, tip and tilt components, we can see additional micrometeoroid impacts as isolated features (see Figure \ref{fig:impact_features} which shows different kinds of impacts). All such events are summarised in the left panel of Figure \ref{fig:meteoroid_and_stability}, which closely matches the recent performance report of the observatory addressing the effects of micrometeoroid impacts \citep{2026Telfer}. The right panel of Figure  \ref{fig:meteoroid_and_stability} shows the difference between two typical consecutive OPD realisations which is a testament to the long term optical stability of the observatory.

\section{Conclusions and Future Work}
\label{sec:conclusions}

\quad The results presented here establish a first open-source approach to model-based wavefront inference with NIRCam, which closely matches but improves upon previous closed-source results. This can now be incorporated as a differentiable module in more complex pipelines. 

We are particularly interested in high-contrast imaging, and below we sketch out next steps towards achieving its potential as a predictive PSF model for calibration-free imaging. \quad It can be hard to find PSF reference stars for high contrast imaging, and imposes significant observing overheads, so a very accurate predictive PSF model would unlock new capabilities. An example is our ongoing \ac{jwst} Program targeting S-stars near the Galactic Centre (GO~11052, PI: S. Ray). Due to the extremely high source density surrounding Sgr~A*, it is impractical to identify isolated calibrators. Data-driven effective PSF models so far reach only moderate contrasts \citep{Calissendorff2023,2023defurio}.

\subsection{Optical Model Improvements}
\label{sec:model_improvements}
\quad Wavefront reconstruction with \camino is a start, but high-contrast imaging will require advances in the optical model, including its amplitude maps and dependence on pointing in time, which will enable calibration-free or telemetry-informed interferometry and coronagraphy.

We currently use a fiducial pupil support model and ignore amplitude variations across it, which may contribute to the residual model mismatch observed in Figure~\ref{fig:model_comparison}, where the $\chi^2_{\nu}$ value remains well above unity. In future work, we hope to allow the pupil geometry and illumination to vary and infer these directly from data.

From ground testing we know there to be a low-order effective wavefront variation with field pointing and between instruments \citep{2016Aronstein,2023PASPRieke,2023PASPMcElwain}. By interpolating defocused observations taken at different positions, and phase retrieval on in-focus sources, we hope to build an accurate data-driven model of this to enable calibrator-free coronagraphy \citep{Feng2025} and interferometry \citep{DeFurio2025calfree}.  

We intend to apply \camino to time domain observations with defocused imaging \citep{2014Beichman} of transits across bright host stars. This has been used not just to measure a transit light curve but also reconstruct a rapid primary mirror segment tilt event that occurred during these observations \citep{2023PASPSchlawin, 2024Telfer}. While this observing mode is not frequently used, applying \camino to archival time series we expect to be able to accurately measure the spatial dependence of the heater mode noted by \citet{2024Telfer}. 

\subsection{Effective Detector Model}

\quad In this paper we have treated pipeline-reduced data, using a linear detector model. The `brighter-fatter effect' and other nonlinear detector crosstalk are a limiting factor in \ac{jwst} high contrast imaging \citep{Sallum2024,Ray2025}, and will print through to wavefront sensing at some level.

In light of this, a next logical step is to build an Effective Detector Model (EDM) as in \cite{2025Desdoigts}, to integrate the detector into the forward modelling process, learning a neural network to predict pixel reads directly from illumination patterns and calibrating this model with a large ensemble of point-source training data.

To build an effective detector model for NIRCam, we would begin with a large ensemble of high signal-to-noise ratio in-focus images that span a wide dynamic range of well depths, with many groups up the ramp. In addition, diversity in terms of wavelength and sub-pixel dither location would further limit the variability in the detector's response and allow clean separation of optical from detector induced effects. Calibration observations made under a consistent optical state (i.e. near-in-time to weak lens wavefront sensor data) would also provide a clear definition for training the detector model. Such an effort would represent a major stride towards developing a predictive PSF model suitable for high contrast applications, where detector based systematic errors often dominate the error budget.

\section{Acknowledgements}

\quad We thank the anonymous referee for their constructive comments and suggestions, which have helped improve this paper.

We acknowledge and pay respect to the traditional owners of the land on which Macquarie University and the University of Sydney are situated, upon whose unceded, sovereign, ancestral lands we work. We pay respects to their Ancestors and descendants, who continue cultural and spiritual connections to Country.

SR, BP, and PT have been supported by the Australian Research Council grant DP230101439; and LD and MC have been supported by the Australian Government Research Training Program (RTP) award. We are grateful to the Australian public for enabling this science. BP and SR would like to thank the Big~Questions~Institute for their philanthropic support. AS, MP, and LP acknowledge support from the NSF, NASA, and the STScI Director’s Discretionary Fund. Development of \dlux has been supported by the Breakthrough Foundation through their Toliman project as a part of the Breakthrough Watch initiative. 

This work is based on observations made with the \\
NASA/ESA/CSA James Webb Space Telescope. The data were obtained from the Mikulski Archive for Space Telescopes at the Space Telescope Science Institute, which is operated by the Association of Universities for Research in Astronomy, Inc., under NASA contract NAS 5-03127 for \ac{jwst}. We are grateful for the support of the Space Telescope Science Institute for their extensive work in understanding, commissioning, and supporting \ac{jwst} and its instruments, including significant correspondence beyond coauthors in this paper and our related works.

Data analysis presented in this work was performed on the OzSTAR national facility at Swinburne University of Technology, which receives funding in part from the Astronomy National Collaborative Research Infrastructure Strategy (NCRIS) allocation provided by the Australian Government, and from the Victorian Higher Education State Investment Fund (VHESIF) provided by the Victorian Government.

This research made use of \textsc{NumPy} \citep{numpy}; \textsc{SciPy} \citep{scipy}; Matplotlib \citep{matplotlib}; \textsc{Jax} \citep{jax}; \numpyro \citep{numpyro_1,numpyro_2}; \equinox \citep{equinox}; \optax \citep{optax2020github}; and ChainConsumer \citep{Hinton2016}.

\bibliography{references}

\end{document}